\documentclass[a4paper,fleqn,usenatbib,amsmath,amssymb]{mnras}

\def\MyApJ#1{}
\def\MyMNRAS#1{#1}
\MyMNRAS{\RequirePackage{rotating}}

\usepackage{graphicx}
\usepackage{dcolumn}
\usepackage{bm}
\usepackage{epsf}
\usepackage{verbatim}
\usepackage{amsmath}
\usepackage{amssymb}
\usepackage{amsfonts}
\usepackage{ifthen}
\usepackage{epsfig}
\usepackage{verbatim}
\usepackage{dcolumn}
\usepackage{epstopdf}
\usepackage{threeparttable}
\usepackage{booktabs}
\usepackage{calc}
\usepackage{multirow}
\usepackage{rotating}
\usepackage{comment}

\usepackage{footnote}
\makesavenoteenv{tabular}

\MyMNRAS{\usepackage{pdflscape}}

\DeclareSymbolFont{CMletters}{OML}{cmm}{m}{it}
\DeclareMathSymbol{v}{\mathord}{CMletters}{`v}

\newcommand{\subfigimg}[3][,]{%
  \setbox1=\hbox{\includegraphics[#1]{#3}}
  \leavevmode\rlap{\usebox1}
  \rlap{\hspace*{10pt}\raisebox{\dimexpr\ht1-2\baselineskip}{#2}}
  \phantom{\usebox1}
}

\def\myfig#1{./Figures/#1}
\def\mybib#1{./#1}

\newcommand{\Mach}{\mathcal{M}}

\newcommand{\MRat}{{f_M}}
\newcommand{\MyR}{\mathcal{R}}
\newcommand{\MyX}{\xi}

\MyApJ{\bibliographystyle{apj}}
\MyMNRAS{\bibliographystyle{mnras}}

\defcitealias{ReissKeshet2014}{RK14}

\defcitealias{KeshetEtAl2010}{K10}

\defcitealias{Keshet2012}{K12}

\newcommand{\ie}{\emph{i.e.} }
\newcommand{\eg}{\emph{e.g.,} }

\newcommand{\be}{\begin{equation}}
\newcommand{\ee}{\end{equation}}
\newcommand{\bea}{\begin{equation*}}
\newcommand{\eea}{\end{equation*}}
\newcommand{\beqr}{\begin{eqnarray} \nonumber}
\newcommand{\eeqr}{\end{eqnarray}}
\newcommand{\beqrb}{\begin{eqnarray}}
\newcommand{\eeqrb}{\nonumber \end{eqnarray}}

\newcommand{\cm}{\mbox{ cm}}

\newcommand{\se}{\mbox{ s}}

\newcommand{\Gyr}{\mbox{ Gyr}}

\newcommand{\kpc}{\mbox{ kpc}}
\newcommand{\Mpc}{\mbox{ Mpc}}

\newcommand{\keV}{\mbox{ keV}}

\usepackage{xcolor}
\definecolor{darkgreen}{rgb}{0.0,0.5,0.0}
\MyApJ{\usepackage[colorlinks,citecolor=darkgreen]{hyperref} 
}

\def\DrawFig#1{#1}

\def\HiRes#1{}
\def\LoRes#1{#1}

\def\FigZSize{9.3cm}
\def\FigASize{8cm}
\def\FigBSize{7.1cm}

\newcommand{\MyTitle}{{Protruding bullet heads indicating dark matter pull}}
\newcommand{\MySTitle}{{Dark matter imprint on bullets}}

\MyApJ{
    
\begin{document}
    \title{\MyTitle}
    \shorttitle{\MySTitle}
	\shortauthors{Keshet et al.}
	\author{Uri Keshet$^{\ddagger}$}
	\author{Itay Raveh}
	\author{Yossi Naor}
	\affil{Physics Department, Ben-Gurion University of the Negev, Be'er-Sheva 84105, Israel}
	\thanks{$^{\ddagger}$Electronic address: ukeshet@bgu.ac.il}
}   	
	
\MyMNRAS{
    \title[\MySTitle]{\MyTitle}
	\author[Keshet et al.]{Uri Keshet\thanks{E-mail: ukeshet@bgu.ac.il}, Itay Raveh, \& Yossi Naor	\\
	Physics Department, Ben-Gurion University of the Negev, POB 653, Be'er-Sheva 84105, Israel}
}

\MyApJ{\date{\today}}

\MyMNRAS{
    \begin{document}
    \pubyear{2020}
    \label{firstpage}
    \pagerange{\pageref{firstpage}--\pageref{lastpage}}
    \maketitle
}

\begin{abstract}
A clump moving through the intracluster medium of a galaxy cluster can drive a bow shock trailed by a bullet-like core.
In some cases, such as in the prototypical Bullet cluster, X-rays show a gas bullet with a protruding head and pronounced shoulders.
We point out that these features, while difficult to explain without dark matter (DM), naturally arise as the head of the slowed-down gas is gravitationally pulled forward toward its unhindered DM counterpart.
X-ray imaging thus provides a unique, robust probe of the offset, collisionless DM, even without gravitational lensing or other auxiliary data.
Numerical simulations and a toy model suggest that the effect is common in major mergers, is often associated with a small bullet-head radius of curvature, and may lead to distinct bullet morphologies, consistent with observations.
\end{abstract}

\date{Accepted ---. Received ---; in original ---}
	
\MyApJ{\label{firstpage}}

\MyMNRAS{
}
	
\MyApJ{	
\maketitle
}
	
\section{Introduction}
\label{sec:Introduction}

Large-scale structure mergers provide a unique laboratory for studying dark matter (DM), in particular as collisions can pull baryons out of their DM halos, leaving the DM and other components spatially separated, with distributions that depend on the dynamics and on interactions.
Offsets between the gas and DM components in such dissociative mergers were measured fairly directly, by mapping the gas using X-ray imaging, and the DM using gravitational lensing.
Indirect evidence from broadband observations supports and supplements the picture; in particular, galaxies tend to coincide with the DM distribution, thus demonstrating the collisionless nature of the latter.
For reviews, see \citet{MarkevitchVikhlinin2007, Molnar15, GolovichEtAl19}.

The prototypical dissociative merger is the galaxy cluster 1E 0657-56, known as the Bullet cluster owing to the distinct bullet-shaped clump moving supersonically through the intracluster medium (ICM) of the host cluster, as inferred from X-ray imaging; see Fig.~\ref{fig:BulletFigs}.
Observations of the gas combined with both weak and strong gravitational lensing indicate a substantial spatial offset between the gas and DM components of the clump, with member galaxies approximately coincident with the DM \citep[][and references therein]{MarkevitchEtAl2002,CloweEtAl2004,Markevitch2006,BradacEtAl2006, DiMascoloEtAl2019}.

Offsets between gas and DM were observed in additional merging galaxy clusters and groups, including
A520 \citep{MahdaviEtAl07},
MACS J0025.4-1222 \citep{BradacEtAl08},
A2163 \citep{OkabeEtAl11},
A1758N \citep{RagozzineEtAl12},
DLSCL J0916.2+2951 \citep['Musket ball cluster';][]{DawsonEtAl12},
MACS J0717.5+3745 \citep{MroczkowskiEtAl12},
SL2S J08544-0121 \citep['Bullet group';][]{GastaldelloEtAl14},
MACS J1149.5+2223 \citep{GolovichEtAl16},
A2034 \citep{MonteiroOliveiraEtAl18},
and others systems which are studied below.
Indirect evidence suggests dissociative mergers in many additional systems, notably in
A1240 \citep{BarrenaEtAl09}, 
CL 0152–1357 \citep{MassardiEtAl10},
ZwCL 0008.8+5215 \citep{vanWeerenEtAl11, MolnarBroadhurst18},
and
A2399 \citep{LourencoEtAl20},
although in some such cases lensing followup shows no DM dissociation \citep[for example in ACT-CL J0102–4915, 'El Gordo';][]{MenanteauEtAl12, JeeEtAl14, DiegoEtAl20}.

While evidence for a DM component separated from the gas in such dissociative mergers is compelling, it is largely based on gravitational lensing juxtaposed against the inferred gas distribution, and so depends on the standard assumptions and limitations of lensing analyses, and on data availability.
It is interesting to examine if evidence for the DM component is encoded directly on the gas distribution, and to explore what insights on the DM can be inferred robustly from observations of the gas, in particular from X-ray imaging.
Stated differently, we wish to determine which features of the observed gas can be robustly modelled with vs. without DM.
Putative prominent gas features in such mergers that cannot be explained with baryonic physics alone would thus provide a unique probe of a dynamically significant DM component, that is necessarily collisionless to a high degree.
Mergers often lead to highly irregular gas morphologies that are difficult to model, so we focus on systems presenting fairly symmetric, bullet-like geometries, with a distinct bullet trailing a shock.

We begin by studying observations of the Bullet cluster in \S\ref{sec:Bullet}, and argue that the protruding head and pronounced concave shoulders of the bullet are not easily explained without invoking DM.
We show how DM naturally leads to the observed morphology of the Bullet cluster in \S\ref{sec:BulletHead}, and to bullets with protruding heads and concave shoulders in general in \S\ref{sec:Model}, using hydrodynamical simulations and a toy model.
Protruding heads are thus expected to be quite common in major mergers, in accordance with their appearance in a fair fraction of the relevant observed systems, and attributing them to the pull of DM is more plausible than alternative scenarios, as we argue in \S\ref{sec:BulletsGeneral}.
Our results are summarized and discussed in \S\ref{sec:Discussion}.
We adopt a $\Lambda$CDM model with a Hubble parameter $H_0=70\,\mbox{km}\,\mbox{s}^{-1}\,\mbox{Mpc}^{-1}$,
a matter fraction $\Omega_m=0.3$, a baryon fraction $f_b\equiv\Omega_b/\Omega_m=0.17$, and a $\Gamma=5/3$ adiabatic index.

\section{Bullet cluster: protruding head}
\label{sec:Bullet}

The Bullet cluster presents one of the hottest, most X-ray luminous ICMs, harboring one of the strongest known merger shocks, with a Mach number $\Mach=3.0\pm0.4$ \citep{MarkevitchEtAl2002, Markevitch2006, DiMascoloEtAl2019}.
The merger plane appears to approximately coincide with the plane of the sky, and the observation shows the merger only $\sim0.15\Gyr$ after nearest approach \citep{MarkevitchEtAl2002, BarrenaEtAl2002,MarkevitchEtAl2002,MastropietroBurkert2008, WittmanEtAl18}.
Observations indicate that the distribution of member galaxies approximately follows the distribution of mass \citep{CloweEtAl2004}, whereas the distributions of gas density \citep{MarkevitchEtAl2002} and pressure \citep[inferred from the Sunyaev-Zel'dovich effect;][]{MaluEtAl10} are offset from the mass and from each other, providing a testbed for DM and baryonic physics \citep[\eg][]{CloweEtAl2004, BradacEtAl2006, MarkevitchVikhlinin2007, LeeKomatsu2010, GrahamEtAl2015}.

Figure \ref{fig:BulletFigs} (top left panel) shows an exposure corrected, background subtracted, $0.8 \mbox{--} 7.0$ keV \emph{Chandra} image of the bullet merger region.
The image was smoothed with a $3''$ Gaussian, where $1''$ is equivalent to a proper projected distance of $\sim4.4\kpc$ at the $z\simeq 0.296$ redshift of the cluster.
Table~\ref{tab:Obsids} lists the observation parameters.
Periods of elevated background are identified using the $2.5\mbox{--}7.0$ keV lightcurve in a background region free of cluster emission on the Advanced CCD Imaging Spectrometer (ACIS) chip.
Standard procedures are used for modelling the detector readout artifacts and sky background, and exposure maps are created using
Alexey Vikhlinin's tools\footnote{\href{http://hea-www.harvard.edu/~alexey/CHAV/}{http://hea-www.harvard.edu/\~{}alexey/CHAV/}}.
These exposure maps account for the position- and energy-dependent variation in effective area and detector efficiency \citep{WeisskopfEtAl2002} using a MEKAL model.
The exposure maps and the images of the cleaned event, background, and readout files of the different ObsIDs are co-added in sky coordinates.
After cleaning and co-addition, intensity images in the $0.8\mbox{--}7.0$ keV energy band are created by subtracting both background and readout files from the processed event files, and dividing the outcome by the co-added exposure map.
Point sources are excluded by a visual inspection.

\MyApJ{\begin{table}[h]}
\MyMNRAS{\begin{table}}
\caption{\label{tab:Obsids}
Bullet-cluster \emph{Chandra} observations used in Fig.~\ref{fig:BulletFigs}.} 
\centering
\setlength{\tabcolsep}{0.5em} 
{
\begin{tabular}{|c|c|c|c|} 
\hline 
ObsID& Start date & Total exposure & Cleaned exposure \\
(1) & (2) &  [ks] (3) & [ks] (4) \\
\hline 
$554$& $2000$-$10$-$16$& $26.8$& $14.5$\\
$3184$& $2002$-$07$-$12$& $88.6$& $78.8$\\
$4984$& $2004$-$08$-$19$& $77.6$& $76.7$\\
$4985$& $2004$-$08$-$23$& $28.9$& $23.8$\\
$4986$& $2004$-$08$-$25$& $42.0$& $40.4$\\
$5355$& $2004$-$08$-$10$& $29.5$& $20.7$\\
$5356$& $2004$-$08$-$11$& $98.9$& $97.5$\\
$5357$& $2004$-$08$-$14$& $80.7$& $79.8$\\
$5358$& $2004$-$08$-$15$& $32.8$& $32.1$\\
$5361$& $2004$-$08$-$17$& $83.9$& $83.0$\\
\hline \end{tabular}}
\vspace{0.1cm}
\begin{tablenotes}
\item Columns: (1) Observation ID number; (2) Observation start date; (3) Total observation exposure; (4) Total exposure after the cleaning process (see text).
\end{tablenotes}
\end{table}

\begin{figure*}
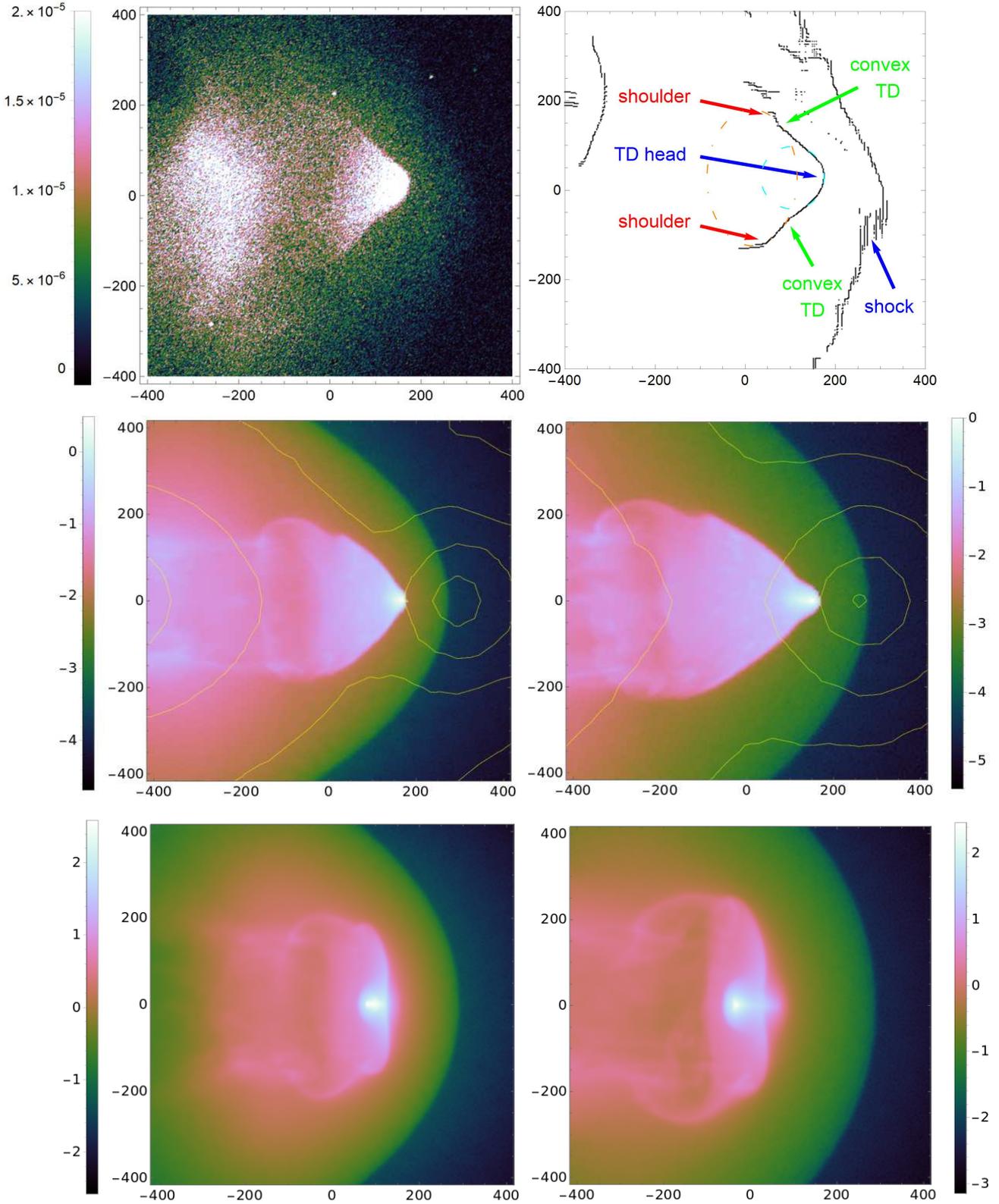
\MyApJ{[h!]}
\centering
\DrawFig{
\hspace{-21mm}
\HiRes{\includegraphics[width=\FigZSize]{\myfig{HalfZoom3LegL.eps}}}
\LoRes{\includegraphics[width=\FigZSize]{\myfig{HalfZoom3LegL_LR800.eps}}}
\raisebox{4mm}{
\hspace{-3mm}
\includegraphics[width=\FigBSize]{\myfig{LabelledBullet5.eps}}
} \\
\vspace{-2mm}
\HiRes{\includegraphics[width=\FigASize]{\myfig{EM_011120_NFWT_DM_VHR_C3_S142.eps}}}
\LoRes{\includegraphics[width=\FigASize]{\myfig{EM_011120_NFWT_DM_VHR_C3_S142_R600.eps}}}
\HiRes{\includegraphics[width=\FigASize]{\myfig{EM_011120_NFWT_DM_VHR_C3_S152.eps}}}
\LoRes{\includegraphics[width=\FigASize]{\myfig{EM_011120_NFWT_DM_VHR_C3_S152_R600.eps}}}
\\
\vspace{-2mm}
\hspace{0.2mm}
\HiRes{\includegraphics[width=\FigASize]{\myfig{EM_021120_NFWT_gas_VHR_C3_S142.eps}}}
\LoRes{\includegraphics[width=\FigASize]{\myfig{EM_021120_NFWT_gas_VHR_C3_S142_R600.eps}}}
\HiRes{\includegraphics[width=\FigASize]{\myfig{EM_021120_NFWT_gas_VHR_C3_S152.eps}}}
\LoRes{\includegraphics[width=\FigASize]{\myfig{EM_021120_NFWT_gas_VHR_C3_S152_R600.eps}}}
}
\caption{
The Bullet cluster in observations (upper row), gas+DM simulations (middle row) and gas-only simulations (bottom row).
Upper left panel: \emph{Chandra} $0.8\mbox{--}7.0\keV$ (colourbar: flux $[\mbox{cnt s}^{-1}\cm^{-2}]$) image of $\sim3'$ width (corresponding to $\sim800\kpc$; ticks) centreed on J2000 $\{\mbox{RA}=6^{\circ}58'24'',\delta=-55^{\circ}56'36''\}$ (west to the right).
Upper right: same image, labelled (see text), after applying a $14''$ Canny edge detector.
Middle row: nominal gas+DM simulation, showing the emission measure for gas ($\log_{10}\mbox{EM}[\mbox{cm}^{-6}\kpc]$ colourbar) and the total surface mass for DM (contours of factor 2 increments) during best alignments of DM peaks ($t=1.42\Gyr$ after simulation start, left) and of the bullet ($t=1.52\Gyr$).
Bottom row: same as middle row, but with DM replaced by gas, at the same simulation times.
}
\label{fig:BulletFigs}
\end{figure*}

As Fig.~\ref{fig:BulletFigs} shows, the projected gas distribution is bimodal, with a dense, bullet-like clump to the west, and a more diffuse clump to the east.
The characteristic bullet and bow shock, oriented approximately toward the west, are evident, and sufficiently pronounced to be picked up by a standard \citet{canny1986computational} edge detector, as shown in the top right panel of the figure.
The resulting edges include the (labelled) tangential discontinuity (TD), outlining the bullet near the centre of the image, and the shock front to its west.
The bullet is thought to carry the west-oriented momentum of the incident minor subcluster, after it collided with the core of the major subcluster and drove the shock.
Gravitational lensing indicates a bimodal mass distribution which is offset from the gas, with a smaller clump to the west approximately coincident with the shock, and a larger clump southeast of the diffuse clump.

As seen in the top-left image and highlighted in the top-right panel, the bullet presents pronounced shoulders and a head protruding to the west.
In terms of the flow around a solid object, the body outlined by the bullet TD is concave at the shoulders.
This differs from the standard bullet TD formed by a moving source, which is typically regular and convex.
The observed TD geometry is rather unexpected to form naturally in a steady flow, as convex regions typically generate additional shocks \citep[\eg][and references therein]{GrishinPogorelov88, ChenFeldman15}, instabilities \citep[\eg][and references therein]{morton1996stagnation,PanarasDrikakis11}, and drag \citep[except in special cases; \eg][]{GusarovLevin82,Plakhov16}.
The evolving flow around the TD is not expected to be steady, especially not so close to nearest passage, but as we show below, gasdynamic effects including instabilities do not naturally explain the observed bullet morphology.
Similar protruding bullet heads and concave shoulders are found in additional observed mergers, reviewed below.
It is therefore interesting to examine if the observed morphology may be a telltale dynamical sign of the gravitational effect of DM.

\section{Bullet cluster: the pull of DM}
\label{sec:BulletHead}

Consider a binary collision leading to the formation of a gas bullet, carrying part of the momentum of the incoming mass $M_2$;
in the Bullet cluster, $M_2$ is the smaller clump, arriving from the east.
The gas component of $M_2$ (denoted gas2) is slowed down by the ram pressure exerted by collisions off the gas particles of $M_1$ (gas1), and thus trails its own, collisionless by assumption, DM component (denoted DM2).
The ram pressure increases as gas2 accelerates toward the high density centre of gas1, but is typically relieved gradually after their nearest approach.
If at this stage the DM2 potential well is not too far ahead and not too fast with respect to the gas bullet, the tip of the bullet can accelerate toward DM2, developing a head (dashed cyan circle in the top-right panel of Fig.~\ref{fig:BulletFigs}) protruding from the main bullet (dot-dashed orange), producing concave shoulders.
Subsequently, depending on the parameters, the bullet head can either trail behind DM2, catch up with it, or overshoot it.

The dynamics of the Bullet cluster were simulated by several groups \citep[][and references therein]{Takizawa05, Takizawa06, SpringelFarrar07, MilosavljevicEtAl07, MastropietroBurkert2008, AkahoriYoshikawa2012, Dawson13, LageFarrar14, WittmanEtAl18}.
We follow \citet{SpringelFarrar07}, who used the $N$-body/smoothed particle hydrodynamics (SPH) code {\scriptsize GADGET2} \citep{SpringelEtAl01, Springel05} to simulate the merger of two bodies, each with overlapping NFW \citep{NavarroEtAl1997} distributions of both DM and gas.
For simplicity, we adopt the same setup: a mass ratio $\MRat\equiv M_1/M_2=10$, head-on, binary collision, with zero energy orbits.
Here, $M$ is defined as the total mass inside the virial radius $\MyR=\MyR_{200}$ of each structure, enclosing a density $200$ times larger than the critical density of the Universe.

In more detail, we adopt $M_{2}=1.5\times 10^{14}M_\odot$ with a virial radius $\MyR_2\simeq 1.1\Mpc$, initially moving westward.
The initial, spherically symmetric distribution of each component $j$ is given by hydrostatic equilibrium with the NFW radial profile of mass density
\begin{equation} \label{eq:NFW}
\rho_j(r) = \frac{\rho_{j0}}{(r/R)(1+r/R)^2} \, ,
\end{equation}
truncated at $r=\MyR=c R$, where $R$ is the scale radius, $c$ is the concentration parameter, and $\rho_{j0}$ is the corresponding normalization.
The gas distributions are initialized with small cores in their centres.
The simulation is carried out in the centre of mass frame, starting with the virial radii of the two bodies touching each other.
Our nominal runs have $1.6\times 10^7$ particles; more than sufficient for convergence, as we confirm by examining higher and lower resolutions.

The middle row of Fig.~\ref{fig:BulletFigs} shows the simulated gas emission measure (colourbar) and DM surface mass (contours) for the nominal concentration parameters $c_1=3$ and $c_2=7.2$, found by \citet{SpringelFarrar07} to reproduce fairly well the bullet morphology, the distance between the bullet and the shock, and the distance between the two mass peaks.
While the setup is simplified and approximate, the morphologies of the simulated shock and bullet qualitatively agree with observations, showing in particular the protruding bullet head extending into the DM potential well, and the resulting concave shoulders.
The bottom row shows the same simulation, but with $f_b=1$, \ie with no DM.
In this case, the bullet has a larger radius of curvature and, as expected in the absence of the DM potential well, does not present a protruding head or concave shoulders.

These results, for both $f_b=0.17$ and $f_b=1$, are robust, and not sensitive to the choice of parameters near their nominal values.
Our simulations thus indicate that a bullet with a protruding head naturally forms in a head-on collision only when its DM counterpart is present, and provided that the ram pressure is not sufficiently high to efficiently detach the bullet from DM2, as we next show.

\section{Bullet and head formation}
\label{sec:Model}

Consider a head-on collision as in \S\ref{sec:BulletHead} between two structures of masses $M_1$ and $M_2<M_1$, characteristic radii $R_1$ and $R_2$, and virial radii $\MyR_1$ and $\MyR_2$.
To accurately model the formation of the bullet and its possibly protruding head, it is necessary to evolve the axisymmetric but non-spherical distributions of DM and gas, derive the non-steady, compressible flow in 3D starting from the specific initial conditions, taking into account the evolving gravitational potential and gasdynamic effects such as ram pressure, the gradual stripping of the gas, the formation of the bow shock and its effect on the flow, and dissipative processes, as well as the violent relaxation of DM2.
Nevertheless, a simplified analytic treatment can be useful \citep[\eg][]{Takizawa06}.

For our purposes, suffice to invoke an even simpler, effectively 1D picture, in which the relative velocity $(v_2-v_1)$ between the two masses is approximately fixed till nearest approach by its initial value as the structures first touch,
\begin{equation}
(v_2-v_1)^2 \simeq \frac{2G M_1 M_2}{(\MyR_1+\MyR_2)\mu} = \frac{2G (M_1+M_2)}{\MyR_1+\MyR_2} \,,
\end{equation}
where $\mu\equiv M_1M_2/(M_1+M_2)$ is the reduced mass and $G$ is the gravitational constant.
By nearest approach, gas2 stripped to its core mass $m_2$ has effectively shared its initial momentum with some gas1 mass $m_1$, and so moves at a velocity slower than DM2 by
\begin{equation}
\Delta v \simeq  v_2 - \frac{m_1 v_1 + m_2 v_2}{m_1+m_2} \simeq \frac{m_1}{m_1+m_2}\sqrt{\frac{2G (M_1+M_2)}{\MyR_1+\MyR_2}} \, .
\end{equation}
One can compare $\Delta v$ to the escape velocity $u_2\simeq (-2\Phi_2)^{1/2}$ from the centre of the DM2 potential $\Phi_2$.
If $\Delta v<\MyX u_2$, where $\MyX$ is a dimensionless factor of order unity, the gas will remain bound to DM2, and a bullet with a protruding head is expected to form as some of the gas accelerates into DM2.
If $\Delta v>\MyX u_2$, the gas escapes from the potential well toward DM1, and even if a bullet forms, it may be too far from DM2 to develop a protruding head.

For simplicity, we approximate $m_2\simeq f_b M_2(r<R_2)$ as the initial mass of the gas2 core inside $R_2$, and $m_1\simeq f_b M_1(\bar{z}<0,\bar{\varrho}<R_2)$ as the initial gas1 mass inside the cylinder carved by the motion of this gas2 core from first contact till nearest approach. Here, we used cylindrical coordinates $\{\bar{\varrho},\bar{\varphi},\bar{z}\}$, with $\bar{z}$ being the displacement from the centre of gas1 along the collision axis.
As Fig.~\ref{fig:Scan} shows, bullets with a protruding head falling onto DM2 (blue up triangles) are well distinguished by their small $\Delta v/u_2$ (ordinate) from bullets that detach from DM2 with no protruding head (red down triangles), for a wide range of $\MRat$, $c_1$, and $c_2$ values.
The factor $\MyX\simeq 0.4\mbox{--}0.6$ depends on all three parameters, but is approximately a function of the ratio $f_P$ between the typical ram pressure $\rho(R)v^2$ exerted by each incoming gas component.
At intermediate, $\Delta v\gtrsim \MyX u_2$ values, some detached bullets (black circles) have sufficient time to develop a protruding head before escaping from DM2.
Explicit expressions for $m_1$, $m_2$, $u_2$, and $f_P$ are provided in \S\ref{sec:appNFW} for the case of NFW distributions.

\begin{figure}\MyApJ{[h!]}
\centering
\DrawFig{
\includegraphics[width=8.5cm]{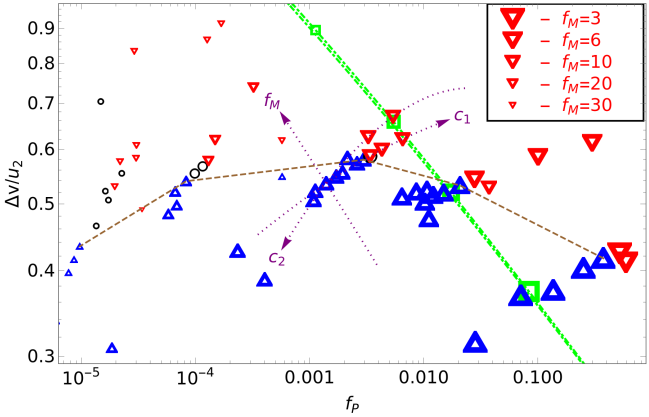}
}
\caption{
Outcome of binary head-on collisions in {\scriptsize GADGET2} simulations for different merger parameters $\MRat$ (symbol size; see legend), $c_1$, and $c_2$ (dotted purple curves demonstrate the variation of each parameter with respect to the nominal case of \S\ref{sec:BulletHead}).
For a small lag-to-escape velocity ratio $\Delta v/u_2$ (blue up triangles), the gas bullet forms a protruding head that falls onto the DM2 potential well.
For large $\Delta v/u_2$, the gas escapes from DM2 before (red down triangles), but sometimes after (black circles) forming a protruding head.
The DM dissociation threshold $\MyX$ is approximately a function (dashed brown curve) of the ram-pressure ratio $f_P$.
The mean mass--concentration relation \citep[\eg][green dot-dashed curves shown for both $z=0$ and $z=1$ redshifts, with squares of sizes indicating $f_M$]{RagagninEtAl19} favors bullets with protruding heads in major mergers (small $\MRat$).
}
\label{fig:Scan}
\end{figure}

While the precise value of $\MyX$ and other details of Fig.~\ref{fig:Scan} depend on the underlying assumptions, the general picture is robust, with a protruding bullet head forming when the merged gas cannot escape DM2.
In particular, conservative modifications of the NFW distributions of gas and DM, for example by switching to \citet{Hernquist90} profiles or avoiding profile truncation, yield qualitatively similar results.
Within the NFW framework, the mean mass--concentration relation $c\propto M^{-0.1}$ \citep[\eg][and references therein]{RagagninEtAl19} suggests that bullets with protruding heads should be quite common in major mergers, as indicated by the green dot-dashed curves and squares in the figure.

\section{Other bullets and models}
\label{sec:BulletsGeneral}

We review the literature in search of pronounced bullets trailing bow shocks in high-quality X-ray imaging data, avoiding highly irregular morphologies (such as in A520) which may indicate a late stage, non-binary, or unfavourably-projected merger.
Eleven systems suggestive of a binary merger close to the plane of the sky are listed in Table \ref{tab:OtherClusters} (see literature references therein).
Among the 11 bullets, six (including the Bullet cluster) show evidence of a protruding head, as observed in the Bullet cluster.
Although lensing data are available for only eight of the 11 systems, it is interesting that a DM clump is observed leading three of the bullets with a protruding head, but none of the bullets without a protruding head, in accordance with our model.

\defcitealias{Andrade-SantosEtAl19}{AS19}
\defcitealias{Botteon16}{B16}
\defcitealias{ColemanEtAl17}{C17}
\defcitealias{DasadiaEtAl16}{D16}
\defcitealias{DiegoEtAl20}{D20}
\defcitealias{EmeryEtAl17}{E17}
\defcitealias{GolovichEtAl17}{G17}
\defcitealias{GolovichEtAl19}{G19}
\defcitealias{HlavacekLarrondoEtAl18}{H18}
\defcitealias{KingEtAl16WL}{K16}
\defcitealias{JeeEtAl14}{J14}
\defcitealias{MacarioEtAl11}{M11}
\defcitealias{MarkevitchEtAl2002}{M02}
\defcitealias{Markevitch2006}{M06}
\defcitealias{MertenEtAl11}{M11b}
\defcitealias{MolnarBroadhurst18}{M18}
\defcitealias{Monteiro-OliveiraEtAl17}{M17}
\defcitealias{OgreanEtAl13}{O13}
\defcitealias{OkabeUmetsu08}{O08}
\defcitealias{OwersEtAl11_A2744}{O11}
\defcitealias{RussellEtAl2012}{R12}
\defcitealias{RussellEtAl14}{R14}
\defcitealias{Urdampilleta18}{U18}
\defcitealias{vanWeerenEtAl11}{W17}
\defcitealias{WhiteEtAl15}{W15}
\begin{table*}\MyApJ{[h!]}
\caption{
\label{tab:OtherClusters}
Galaxy clusters showing an apparent binary merger with a bullet trailing a shock in high-resolution X-ray imaging.
}
\begin{center}
\begin{tabular}{|c|c|c|c|c|c|c|}
\hline
Cluster name (bullet orientation) & Leading DM  & Protruding & Shock Mach & Expected & Measured & References \\
(1) & offset (2) & head (3) & $\Mach$ (4) & $\alpha$ (5) & $\alpha$ (6) & (7) \\
\hline
\hspace{-0.2cm} ACT-CL J0102–4915 (El Gordo; SE) \hspace{-0.2cm} & no (WL+SL) & no  & $\gtrsim 3$ 
& $\lesssim0.3$ & $\sim 1.5$ & \citetalias{JeeEtAl14, Botteon16, DiegoEtAl20} \\ 
A3376 (E) & no (WL) & no & $1.5\pm0.1$ & $0.7\pm0.2$ & $\sim1$ & \citetalias{Monteiro-OliveiraEtAl17, Urdampilleta18} \\ 
A754 (SE) & no$^\dagger$ (WL) & no & $1.57_{-0.12}^{+0.16}$ & $0.7_{-0.2}^{+0.1}$ & $\sim 0.5$ & \citetalias{OkabeUmetsu08, MacarioEtAl11} \\
ZwCL 0008.8+5215 (W) & no (WL+RS) & no & $2.4_{-0.2}^{+0.4}$ (R) & $0.3\pm0.1$ & $\sim2$ (R) & \citetalias{vanWeerenEtAl11, MolnarBroadhurst18, GolovichEtAl17} \\
Cluster around 3C 438 (SE) & --- & no & $2.3\pm0.5$ & $0.3\pm0.2$ & $\sim 0.2$ & \citetalias{EmeryEtAl17} \\
1E 0657-56 (Bullet, W) & yes (WL+SL+RS) & yes & $3.0\pm0.4$ & $0.3\pm0.1$ & $\sim 2$ & \citetalias{MarkevitchEtAl2002,Markevitch2006} \\
A2146 (SE) & no$^\ddagger$ (WL+SL) &  elongated & $2.3\pm0.2$ & $0.3\pm0.1$ & $\sim 4$ & \citetalias{RussellEtAl2012, KingEtAl16WL, ColemanEtAl17} \\
A2744 (SE) & yes (WL+SL) & elongated & $1.41_{-0.08}^{+0.11}$ & $0.9\pm0.2$ & $\sim 4$ & \citetalias{OwersEtAl11_A2744, MertenEtAl11} \\
RX J0334.2-0111 (NE) & --- &  elongated & $1.6_{-0.3}^{+0.5}$ & $0.6_{-0.2}^{+0.6}$ & $\sim 4$ & \citetalias{DasadiaEtAl16} \\
RX J0751.3+5012 (NW) & --- & yes & $1.9\pm0.4$ & $0.4_{-0.1}^{+0.3}$ & $\sim 0.3$ & \citetalias{RussellEtAl14}\\
1RXS J0603 (S) 
& yes (RS) & yes & $1.5\pm0.1$ & $0.7\pm0.2$ & $\sim1$ & \citetalias{OgreanEtAl13, GolovichEtAl19}\\
\hline
\end{tabular}
\vspace{0.1cm}
\begin{tablenotes}
Columns:
(1) -- Cluster name (second name and abbrev. bullet orientation in parenthesis);
(2) -- DM offset based on weak lensing (WL), strong lensing (SL), or red sequence galaxies (RS);
(3) -- The appearence of a protruding bullet head;
(4) -- Shock Mach number $\Mach$ based on X-rays (R -- based on radio relic);
(5) -- Shock standoff-to-head curvature ratio $\alpha\equiv d_s/r_h$, computed based on $\Mach$ in the \citet{KeshetNaor2016} approximation;
(6) -- Measured $\alpha$;
(7) -- References:
\citetalias{Andrade-SantosEtAl19} -- \citet{Andrade-SantosEtAl19};
\citetalias{Botteon16} -- \citet{Botteon16};
\citetalias{ColemanEtAl17} -- \citet{ColemanEtAl17};
\citetalias{DasadiaEtAl16} -- \citet{DasadiaEtAl16};
\citetalias{DiegoEtAl20} -- \citet{DiegoEtAl20};
\citetalias{EmeryEtAl17} -- \citet{EmeryEtAl17};
\citetalias{GolovichEtAl17} -- \citet{GolovichEtAl17};
\citetalias{GolovichEtAl19} -- \citet{GolovichEtAl19};
\citetalias{JeeEtAl14} -- \citet{JeeEtAl14};
\citetalias{KingEtAl16WL} -- \citet{KingEtAl16WL};
\citetalias{MacarioEtAl11} -- \citet{MacarioEtAl11};
\citetalias{MarkevitchEtAl2002} -- \citet{MarkevitchEtAl2002};
\citetalias{Markevitch2006} -- \citet{Markevitch2006};
\citetalias{MertenEtAl11} -- \citet{MertenEtAl11};
\citetalias{MolnarBroadhurst18} -- \citet{MolnarBroadhurst18};
\citetalias{Monteiro-OliveiraEtAl17} -- \citet{Monteiro-OliveiraEtAl17};
\citetalias{OgreanEtAl13} -- \citet{OgreanEtAl13};
\citetalias{OkabeUmetsu08} -- \citet{OkabeUmetsu08};
\citetalias{OwersEtAl11_A2744} -- \citet{OwersEtAl11_A2744};
\citetalias{RussellEtAl14} -- \citet{RussellEtAl14};
\citetalias{RussellEtAl2012} -- \citet{RussellEtAl2012};
\citetalias{Urdampilleta18} -- \citet{Urdampilleta18};
\citetalias{vanWeerenEtAl11} -- \citet{vanWeerenEtAl11}. \\
$^\dagger$ - Although the eastern mass appears split to small parts, one of which leads the bullet.\\
$^\ddagger$ - The DM clump appears to somewhat lag the bullet.
\end{tablenotes}
\end{center}
\end{table*}

The observed morphologies of bullets and bullet heads vary among merger systems, consistent with the variability we find in simulations as a function of merger parameters and time, as illustrated in Fig.~\ref{fig:IrregularBullets}.
In some cases, like in RX J0334.2-0111, the protruding head is narrow and elongated, consistent with simulations (an example is shown in panel $a$) in which only a small core managed to keep up with DM2.
In other cases, like in A2146, the head is quite spherical, consistent with simulations (panel $b$) in which the gas has already partly settled in the bottom of the DM2 potential well.
Some bullets, like in ZwCL 0008.8+5215, present as a rounded cone, resembling simulations in which the head just catched up with DM2 (panel $c$).
For some parameters, after the gas reaches DM2 it overshoots it, producing an irregular morphology (panel $d$) which could perhaps explain some observed irregular structures, for example the 'foot' in A520 \citep{WangEtAl2016}.

\begin{figure*}
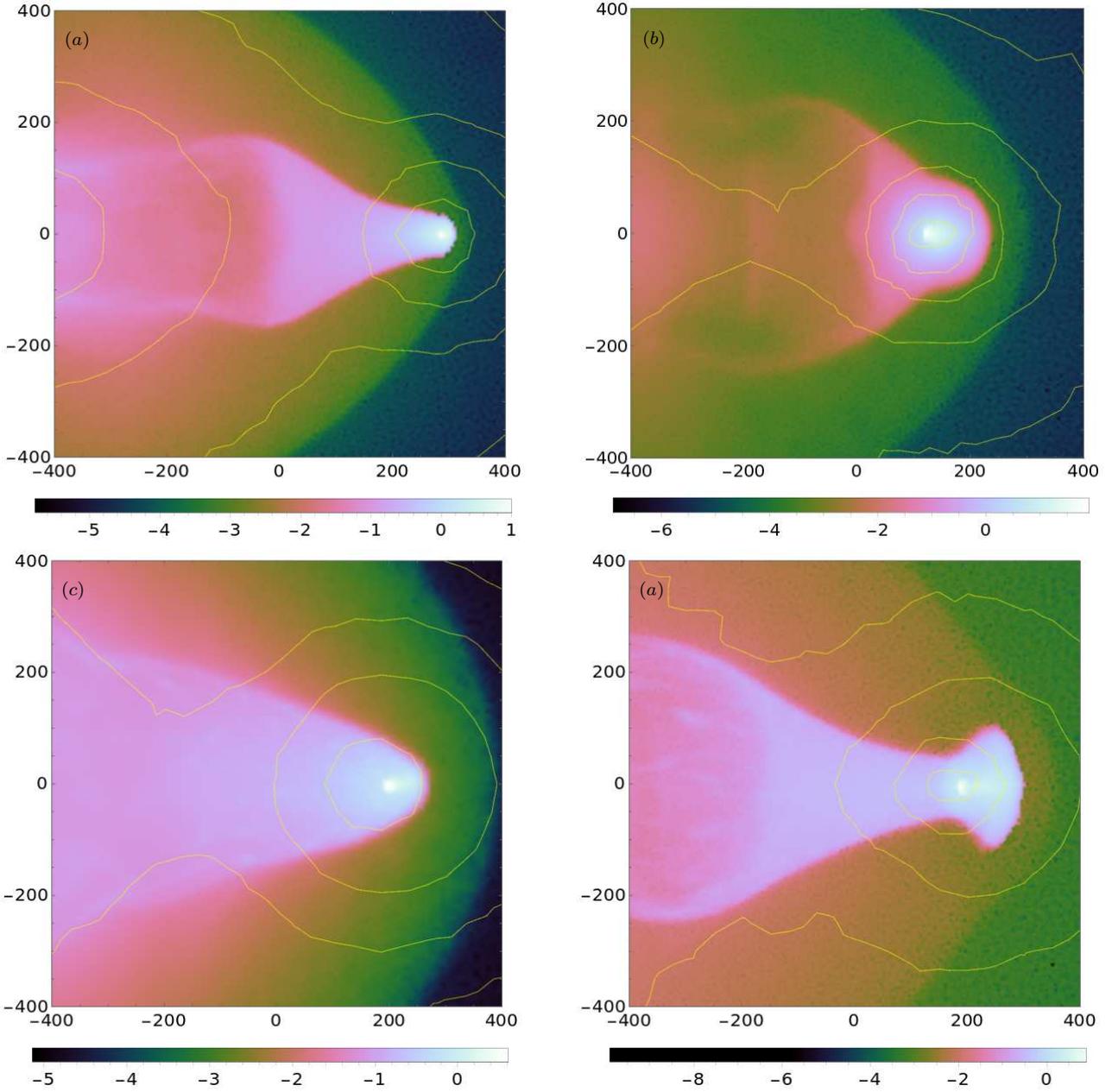
\MyApJ{[h!]}
\centering
\DrawFig{
\HiRes{\subfigimg[width=8cm]{$\quad\quad(a)$}{\myfig{NFWT_DM_HR_C3P12_S140.eps}}}
\LoRes{\subfigimg[width=8cm]{$\quad\quad(a)$}{\myfig{NFWT_DM_HR_C3P12_S140_R600.eps}}}
\hspace{0.5cm}
\HiRes{\subfigimg[width=8cm]{$\quad\quad(b)$}{\myfig{291020_NFW_DM_HR_C3P15_S150.eps}}}
\LoRes{\subfigimg[width=8cm]{$\quad\quad(b)$}{\myfig{291020_NFW_DM_HR_C3P15_S150_Rast.eps}}}
\\
\HiRes{\subfigimg[width=8cm]{$\quad\quad(c)$}{\myfig{NFWT_DM_HR_rM3_C3.25P6_S160.eps}}}
\LoRes{\subfigimg[width=8cm]{$\quad\quad(c)$}{\myfig{NFWT_DM_HR_rM3_C3.25P6_S160_R600.eps}}}
\hspace{0.5cm}
\HiRes{\subfigimg[width=8cm]{$\quad\quad(a)$}{\myfig{NFWT_DM_HR_C3P12_S180.eps}}}
\LoRes{\subfigimg[width=8cm]{$\quad\quad(a)$}{\myfig{NFWT_DM_HR_C3P12_S180_R600.eps}}}
}
\caption{
Different morphologies of simulated bullets, demonstrating a head that is
elongated (panel $a$: $\MRat=10, c_1=3, c_2=12, t=1.40\Gyr$),
nearly spherical (panel $b$: $\MRat=10, c_1=3, c_2=15, t=1.50\Gyr$),
conical (panel $c$: $\MRat=3, c_1=3.25, c_2=6, t=1.60\Gyr$),
or substantially overshooting DM2 (panel $d$: $\MRat=10, c_1=3, c_2=12, t=1.80\Gyr$).
Notations are the same as in Fig.~\ref{fig:BulletFigs} (middle row).
}
\label{fig:IrregularBullets}
\end{figure*}

The deviation of observed bullets from the standard, non-DM, regular and convex morphology has been noted previously, and generally attributed to gasdynamic  effects.
In particular, a combination of Rayleigh-Taylor (RT) and Kelvin-Helmholtz (KH) instabilities was suggested as a possible explanation for the observed morphology of the Bullet cluster, even in the absence of dynamical DM effects, but this appears to hold only at a late stage when the bullet settles and oscillates at the bottom of the potential well \citep[see 'sloshing model' in][]{Takizawa05, AscasibarMarkevitch2006}, unsupported by other observations.
The shoulders in RX J0751.3+5012 were previously interpreted as KH 'wings' \citep{RussellEtAl14}, but KH modes suspected in observations \citep[\eg][]{Machacek06, KraftEtAl11, RoedigerEtAl12a} and reproduced in simulations \citep[\eg][]{Takizawa05, RoedigerEtAl12b} are less regular, massive, and symmetric than seen in the aforementioned merger bullets.

There are additional challenges to the putative interpretation of bullet shoulders as arising from KH instabilities.
Such instabilities do not naturally explain the shoulders of bullets with elongated heads, observed in three of the mergers in Table \ref{tab:OtherClusters} (and further discussed below).
KH instabilities may well be quenched near the heads of bullets by the incoming magnetic field \citep[][and references therein]{VikhlininEtAl01b, RussellEtAl14}, especially as magnetic draping \citep[][and referenced therein]{AsaiEtAl05, Lyutikov2006, DursiPfrommer2008, NaorKeshet15, KeshetNaor2016} substantially strengthens the shock-amplified fields near the TD.
Finally, in contrast to the DM pull on the head of the bullet, KH modes do not diminish the radius of curvature $r_h$ at the tip of the bullet \citep[\eg][]{Takizawa05}, typically quantified in terms of the ratio $\alpha\equiv d_s/r_h$.
Here, $d_s$ is the standoff distance between the shock and the bullet head along the symmetry axis $\bar{z}$.

Indeed, large relative standoff distances $\alpha$ were noticed in several merger bullets, in particular for the elongated head in RX J0334.2-0111, and attributed to the stripping of gas \citep{DasadiaEtAl16} or to other effects, such as the pull of the main cluster, non-sphericity, and projection effects \citep{ZhangEtAl19}.
The ratio $\alpha$ is provided for the above merger systems in Table \ref{tab:OtherClusters}, computed based on the shock Mach number inferred from the X-ray discontinuity (or, in the absence of good X-ray data in ZwCL 0008.8+5215, from the radio relic) using a hodograph-like expansion \citep{KeshetNaor2016, WallersteinKeshet20}, in which the axial flow and shock curvature are simultaneously derived.
\citep[Note that previous studies often invoked approximations such as by][where  unjustified, ad-hoc assumptions on the flow and the shock geometry lead to inaccurate results.]{moeckel1949approximate}

As the table shows, a large $\alpha$ appears to be preferentially associated with a protruding head, especially when elongated.
This disfavors a model based on the stripping of gas, which does not naturally account for the emergence of shoulders; note that gas stripping too can be suppressed by the draped magnetic fields.
Evolutionary effects may indeed modify $\alpha$ with respect to its steady-flow estimate.
However, the typical projected effect we find in our nearly head-on collision simulations is, in the absence of DM, a reduction in $\alpha$, as demonstrated in Fig.~\ref{fig:BulletFigs} (bottom row).
We conclude that the gravitational pull of a leading DM component is the most natural interpretation for the observed protruding heads of bullets, especially when shoulders are pronounced, the head is elongated, or $\alpha$ is large.

\section{Summary and Discussion}\label{sec:Discussion}

A clump moving through the ICM often presents in X-ray imaging as a bullet trailing a bow shock, suggesting a merger close to the plane of the sky.
A fair fraction of the well-observed bullets (see Table \ref{tab:Obsids} and the top row of Fig.~\ref{fig:BulletFigs}) show a protruding head and pronounced, concave shoulders, in some cases with a head radius of curvature smaller than expected in a steady state, gasdynamic picture (\ie a large $\alpha$).
We argue that such features naturally arise due to the gravitational pull of a DM clump leading the bullet, whereas a gasdynamic model is less plausible even when taking into account the non-steady flow, gas stripping, and KH instabilities.
Hydrodynamical simulations produce bullet morphologies (middle row of Fig.~\ref{fig:BulletFigs}; Fig.~\ref{fig:IrregularBullets}) consistent with observations, and calibrate our toy model (\S\ref{sec:Model}), suggesting the prevalence of protruding bullet heads trailing detached DM clumps in major mergers (Fig.~\ref{fig:Scan}).

Protruding bullet heads thus provide a unique, robust probe of the detached, collisionless DM component based on X-ray imaging alone, even without gravitational lensing or other auxiliary data, and independent of the underlying assumptions and limitations thereof.
Such a bullet head provides a visual gauge of the acceleration at the tip of the bullet, pointing in the direction of the potential well.
The different head morphologies demonstrated in Fig.~\ref{fig:IrregularBullets} suggest that additional valuable information concerning an observed merger can be extracted by carefully modeling the head structure, although this would require a more comprehensive study of projected, oblique collisions in 3D, with more realistic gas distributions.

For simplicity, our study focused on head-on binary collisions, and mostly used NFW distributions to represent both DM and gas.
The general conclusions are nevertheless quite robust, with qualitatively similar results obtained for different gas distributions (NFW and Hernquist distributions, with different cutoff radii), a wide range of collision parameters, a small collision obliquity, and other changes in the setup.
For example, while a protruding bullet head requires the presence of the DM2 clump leading the bullet, it does not necessitate a second, DM1 component, and so can be produced in simulated binary collisions where only the small clump has a DM component.

To illustrate the formation of the bullet head, note that its inferred acceleration in the Bullet cluster, $\gtrsim 2\Delta r/\Delta t^2\simeq 6\times 10^{-8}(\Delta r/100\kpc)(\Delta t/0.1\Gyr)^{-2}\cm\se^{-2}$, is not in the deep MOND \citep[modified Newtonian dynamics;][]{Milgrom83} regime, so the presence of a dominant dark mass component pulling the gas should persist also in MOND theory.
Here, the projected head length $\Delta r$ is measured directly, and an upper limit $\Delta t$ on its age can be derived from the bullet size and shock Mach number.

\MyApJ{\acknowledgements}
\MyMNRAS{\section*{Acknowledgements}}
We are greatly indebted to Maxim Markevitch for his extensive help, support, and hospitality.
We thank Ilya Gurwich and Santanu Mondal for helpful discussions.
This research was supported by the Israel Science Foundation (Grant No. 1769/15), by the IAEC-UPBC joint research foundation (Grant No. 300/18), and by the Ministry of Science, Technology \& Space, Israel, and has received funding from the GIF (Grant No. I-1362-303.7/2016).


\bibliography{\mybib{Clusters}}

\MyMNRAS{\label{lastpage}}

\appendix

\section{Dissociation model parameters for NFW distributions}
\label{sec:appNFW}

For the case of NFW distributions, the definitions in \S\ref{sec:Model} yield
\begin{align}
\frac{2m_1 C_1}{f_b M_1} = &
\ln\left[ \frac{(c_1+1)f_R}{\sqrt{f_R^2+c_1^2}+c_1}\right] \\
& + \frac{1}{2F_R}\ln \left[ \left(\frac{F_R-c_1}{F_R+c_1}\right) \left(
\frac{F_R\sqrt{f_R^2+c_1^2}+c_1}{F_R\sqrt{f_R^2+c_1^2}-c_1}
\right) \right]
 \, , \nonumber
\end{align}
\begin{equation}
\frac{m_2 C_2}{f_b M_2} = \ln(2)-1/2 \, ,
\end{equation}
\begin{equation}
u_2^2 = (1-f_b)\frac{2G M_2}{C_2 R_2} \, ,
\end{equation}
and
\begin{equation}
f_P = \frac{C_2}{C_1}
\left(\frac{c_1}{c_2}\right)^3 \MRat^{-2} \, ,
\end{equation}
where we defined $C\equiv \ln(c+1)-c/(c+1)$, $f_R\equiv R_2/R_1$, and $F_R\equiv \sqrt{1-f_R^2}$.

\end{document}